\documentclass{article}

\usepackage{amsmath}
\usepackage{newalg}
\usepackage{graphicx}
\usepackage{url}
\usepackage{wasysym}
\usepackage{float}
\usepackage{hyperref}
\usepackage{natbib}

\newcommand{\pr}{\text{Pr}}
\renewcommand{\vec}[1]{\mathbf{#1}}

\begin{document}

\title{Significance tests for comparing digital gene expression
  profiles}

\author{Leonardo Varuzza\,$^{1,*}$, Arthur Gruber\,$^{2}$ and Carlos
  A. de B. Pereira\,$^{1,}$\footnote{To whom correspondence should be
    addressed.}}

\maketitle

\begin{center}
{\small  $^{1}$Institute of Mathematics and Statistics, Rua do Mat\~{a}o
  1010, Bloco A and $^{2}$Institute of Biomedical Sciences, University
  of S\~{a}o Paulo, 05508-090, S\~{a}o Paulo SP, Brazil}
\end{center}

\begin{abstract}

\section{Motivation:}
Most of the statistical tests currently used to detect differentially
expressed genes are based on asymptotic results, and perform poorly
for low expression tags. Another problem is the common use of a single
canonical cutoff for the significance level (\emph{p-value}) of all
the tags, without taking into consideration the type II error and the
highly variable character of the sample size of the tags.

\section{Results:}
This work reports the development of two significance tests for the
comparison of digital expression profiles, based on frequentist and
Bayesian points of view, respectively. Both tests are exact, and do
not use any asymptotic considerations, thus producing more correct
results for low frequency tags than the $\chi^2$ test. The frequentist
test uses a tag-customized critical level which minimizes a linear
combination of type I and type II errors.  A comparison of the
Bayesian and the frequentist tests revealed that they are linked by a
Beta distribution function. These tests can be used alone or in
conjunction, and represent an improvement over the currently available
methods for comparing digital profiles.

\section{Availability:}
Implementations of both tests are available under the GNU General
Public License at \url{http://code.google.com/p/kempbasu}

\section{Contact:} 
\href{varuzza@gmail.com}{varuzza@gmail.com (L. Varuzza)} and
\href{cpereira@ime.usp.br}{cpereira@ime.usp.br (C.A.B. Pereira)}

\end{abstract}

\section{Introduction}
\label{sec:introduction}

Crucial events in the biology of living organisms, such as cell
differentiation and specialization, depend on minute variations of
gene expression under different conditions and/or temporal events. A
key approach to elucidate a gene function is to quantify and compare
the expression level of a large set of genes in different tissues or
developmental stages, or under different conditions/treatments. This
task can be performed using large-scale hybridization to microarrays,
or by counting gene tags or signatures using methods such as Serial
Analysis of Gene Expression (SAGE), developed by \cite{sage}, and
Massively Parallel Signature Sequencing (MPSS), described by
\cite{mpss}.  By comparing transcript expression profiles among
different samples, one can identify differentially expressed genes
associated with a particular tissue and/or condition. Unlike
microarrays analysis, SAGE and MPSS do not require any prior knowledge
of the transcript sequences. These techniques provide a digital
profiling, and permit to estimate the relative abundance of mRNA
molecules of a transcriptome based upon two main premises. First,
these methods assume that each position-specific short sequence tag
can unequivocally identify its corresponding transcript. Second, that
the tag counts are representative of the abundances of the
corresponding mRNAs of the transcriptome, that is, every mRNA copy has
the same chance of being counted as the corresponding tag of the
library. The selection of a specific tag sequence from the total pool
of transcripts can be well approximated as a sampling with replacement
\citep{Stollberg2000}.

Several statistical tests have been devised to deal with the problem
of comparing digital expression profiles and identifying
differentially expressed genes
\citep{audic1997,baggerly2004,robinson2007,stekel2000,thygesen2006,
  vencio2004}, and reviewed by \cite{man2000}, \cite{romualdi2001} and
\cite{ruijter2002}.  Most of these tests, including the $\chi^2$ test
as the most classical representative, rely on asymptotic methods and,
as such, perform poorly when the sample size is small, as is the case
of low expression tags. Another problematic aspect of detecting
differentially expressed genes by significance tests, is the use of a
single critical level, such as the canonical values 0.1, 0.05 and
0.01, which, apart from the common use, do no present any particular
advantage. In this work we report the development of two distinct
exact significance tests for comparative studies of digital expression
profiles. The methods are based on the frequentist and Bayesian points
of view, respectively, and are fully implemented on open source
programs. Since these significance tests do not use any asymptotic
considerations, they produce more correct results for low frequency
tags than asymptotic methods. Also, the frequentist test uses a
tag-customized critical level which minimizes a linear combination of
type I and type II errors. We provide evidences that both methods
present very similar results and represent an improvement over the
currently available statistical tests.

\section{Methods}

\subsection{Frequentist Significance Test: \emph{p-value}}
\label{sec:kemp}

As a first approach to address the problem of detecting differentially
expressed tags, we propose an exact and novel frequentist significance
test. Assuming $k$ ($>1$) libraries, let $M$ be the number of distinct
tags and $N_j$ the total number of tags in library $j$
($j=1,2,\cdots,k$).  The frequency of the $i$-th tag in the $j$-th
library is denoted by $X_{ij}$. Hence, $N_j =
X_{1j}+X_{2j}+...+X_{Mj}$.  The basic statistical model can be stated
as:

\begin{itemize}
\item For each $j$, the random vector $\vec X_{\bullet j} =
  (X_{1j};X_{2j};\cdots;X_{Mj})$ is distributed as a multinomial with
  parameters $N_j$ and $\vec P_j= (p_{1j};p_{2j};\cdots;p_{Mj})$.
\item The random vectors $\vec X_{\bullet 1}, \vec X_{\bullet
    2},\cdots,\vec X_{\bullet k}$ are mutually statistically
  independent, that is, we assume that the libraries have been
   collected independently.
\item Considering that, for digital expression profiles, parameters
  $\vec P_j$ and $N_J$ usually assume low and high values,
  respectively, several authors
  \citep{audic1997,poisson_sage,Zhu2008} have considered that the
  distribution of tag frequency $X_{ij}$ can be approximated to a
  Poisson with mean $N_jp_{ij}$.
\item Assuming that the libraries have been collected independently,
  the full model for a single tag $i$ is
  \begin{multline}
    \label{eq:full-model}
    \pr\left\{X_{i1}=x_{i1},\cdots, X_{ik}=x_{ik}|
    p_{i1},\cdots,p_{ik}\right\}=\\
    \frac{ (N_1 p_{i1})^{x_{i1}} \cdots (N_k p_{ik})^{x_{ik}}}
    {x_{i1}!\cdots x_{ik}!}
    \exp(N_1 p_{i1}+\cdots+N_k p_{ik}).
  \end{multline}
\end{itemize}

\subsubsection{Partial Likelihood}
\label{sec:partial-likelihood}
One can now write the equation (\ref{eq:full-model}) using an
alternative parametrization.  Let the new parameters be $\theta_i =
N_1p_{i1}+\cdots+N_kp_{ik}$ and $\Pi_i=( \pi_{i1};\cdots; \pi_{ik})$,
with $\pi_{ij} = N_jp_{ij}/\theta_i$. Assuming the random variable
$Y_i = X_{i1}+\cdots+X_{ik}$ and the observation $y_i =
x_{i1}+\cdots+x_{ik}$, the two following events are equivalent:
\begin{gather*}
  \left\{X_{i1}=x_{i1},\cdots,X_{ik}=x_{ik}\right\} \\ 
  \equiv \\
  \left\{ X_{i1}=x_{i1},\cdots,X_{ik}=x_{ik};Y_i = y_i \right\}.  
\end{gather*}

Therefore, the alternative statistical model can be written as:
\begin{multline}
  \label{eq:alt-model}
\pr\left\{X_{i1}=x_{i1},\cdots,
  X_{ik}=x_{ik},Y_i=y_i|\Pi_i,\theta_i\right\}=\\
  \frac{y!}
  {x_{i1}!\cdots x_{ik}!}
  (\pi_{i1})^{x_{i1}} \cdots (\pi_{ik})^{x_{ik}}
  \frac{\theta_i^y e^{-y}}{y!}.
\end{multline}

With this new parametrization, the full likelihood becomes the product
of a multinomial probability function by a Poisson probability
function.  It is noteworthy that the new parameters, $\Pi_i$ and
$\theta_i$, are variation independent, i.e., their values carry no
information about each other.  According to \cite{Basu:2007p418} and
\cite{cox1975pl}, to perform inference about $\Pi_i$ (or $\theta_i$)
one only has to consider as the likelihood the multinomial (or
Poisson) factor of equation (\ref{eq:alt-model}).

Under the full likelihood model, the null hypothesis --  tag
$i$ has the same expression in all libraries -- is

\begin{equation}
  \label{eq:H0}
  H_0: P_i=(p_{i1},\cdots,p_{ik})=(p_i,\cdots,p_i).
\end{equation}

With the new parametrization, the null hypothesis is reduced to a
simple hypothesis as follows:
\begin{equation}
  \label{eq:H0prime}
  H_0': \Pi_i=\Pi_0=\left(\frac{N_1}{N},\cdots,\frac{N_k}{N}\right).
\end{equation}

This approach of partial likelihood, introduced by \cite{cox1975pl},
simplifies considerably the problem of comparing the expression of the
$j$-th tag in all $k$ libraries.  Hence, under the null hypothesis,
the likelihood is simply a multinomial probability function evaluated
for $\Pi_0$.  Being $X_{i\bullet} = (X_{i1},\cdots,X_{ik})$, the
distribution under the null and alternative hypotheses are, respectively,

\begin{multline}
  \label{eq:H0-likelihood}
  H_0': \pr\{X_{i\bullet}|Y_i=y_i;\Pi_i=\Pi_0\} = 
  \frac{y_i!}{N^{y_i}} \prod_{j=1}^k \frac{N_j^{x_j}}{x_j!}
\end{multline}
and
\begin{equation}
  \label{eq:H1-likelihood}
  H_1': \pr\{X_{i\bullet}|Y_i=y_i;\Pi_i\} = 
  y_i! \prod_{j=1}^k \frac{\pi_{ij}^{x_j}}{x_j!}.
\end{equation}

\subsubsection{Significance Level}
\label{sec:p-value}

According to \cite{cox1977rst} and \cite{kempthorne1976uts}, a
significance test is a method that measures the consistency of the
data with the null hypothesis.  The commonly used index to perform
this task is the well known \emph{p-value}.  We refer to
\cite{kempthorne1971psa} for important discussions on the evaluation
of \emph{p-values}.  For a random vector $\vec X$, let $T(\vec X) = T$
be a statistic in which small values of T cast doubt about $H_0$.  For
an observation $\vec x$ with $T(\vec x) = t$, the associated
\emph{p-value} is the probability under $H_0$ of the event $\{T \leq
t\}$, that is, $p = \pr\{T \leq t|H_0\}$. The consequence of this
definition is that the random variable $T$ must be a function that
produces an order in the sample space.  In this ordered sample space,
the sample points with low order favor the alternative hypothesis,
whereas those with a higher order support the null hypothesis.

As we will show below, the likelihood ratio is an appropriate
statistic for ordering the sample space relative to the null
hypothesis.  Let $R(\vec X)$ be the maximum of the likelihood function
under $H_0$ divided by the overall maximum of the likelihood
function (eq. \ref{eq:R}), and $R(\vec x) = r$ be the value of that
statistic at the observation $\vec x$, the \emph{p-value} is $\pr\{R
\leq r|H_0\}$.

\begin{equation}
  \label{eq:R}
  R_i(\vec X)=\left(\frac{y_i}{N}\right)^{y_i}\prod_{j=1}^k \left(\frac{N_j}{x_j}\right)^{x_j}
\end{equation}

As one can conclude, $R$ has the desired property of ordering the sample
space according to the support of $H_0$. The use of likelihood ratios
for computing \emph{p-values} has been already addressed by
\cite{neyman1928}, \cite{pereira1993}, and \cite{dempster1997}.

Let's define the tail set, $T_i$, of frequencies more extreme than
$\vec x_i$: $$T_i = \left\{\vec w =
  (w_1,\cdots,w_k)|w_1+\cdots+w_k=y_i \vee R(\vec w) \leq R(\vec
  x_i)\right\}.$$ The \emph{p-value} of the tag $i$, $pv_i$, that
provides the significance level for $H_0’$ when the observation is
$\vec x_i$, is:
\begin{equation}
  \label{eq:pvalue}
  pv_i=\sum_{w\in T_i} \frac{y_i!}{N^{y_i}}\prod_{j=1}^k \frac{N_j^{w_j}}{w_j!}.
\end{equation}

A serious limitation of this exact \emph{p-value} calculation is that
the number of points in the sample space grows exponentially in regard
to the number of dimensions. To overcome this problem, we use an
algorithm based on the Monte Carlo
method: 
\vspace{0.5cm}

\begin{algorithm}{p-value}{\vec X,\vec N,runs}
  t \= R(\vec X,\vec N) \\
  y \= \sum X_i \\
  \vec p \= \vec N/{\sum N_i} \\
  c \= 0 \\
  \begin{FOR}{i \=1 \TO runs} 
    \vec W \= \text{Random vector with } \mathrm{Multi}(y,\vec p) \\
    \begin{IF}{ R(\vec W,\vec N) \leq t}
      c \= c+1
    \end{IF}
  \end{FOR} \\
  \RETURN c/runs
\end{algorithm}  

In order to test and validate this method, we developed Kemp, a C
language program named after Prof. Oscar Kempthorne, an English
statistician who has produced an extensive work on the topic of
significance tests.

\subsubsection{Critical significance level}
\label{sec:critical-level}

In order to have inferential meaning, the evaluation of a significance
index should help one to decide in favor or against a null hypothesis.
Hence, a decision rule must be stated.  For instance, the critical
level is the cutoff between reject/accept actions. In any digital
expression profile, the relative abundance can vary dramatically from
tag to tag, implying that the use of a single critical significance
level for all tags may be unfair for those tags with low
frequencies. For this reason, we decided to calculate a critical level
for each particular tag, according to the recommendations of
\cite{degroot1986}. Thus, we used the optimum procedure of the
decision theory, which minimizes the risk function $a\alpha + b\beta$,
a linear combination of the two kinds of errors: $\alpha$ and $\beta$,
corresponding to errors of type I (false positive) and type II (false
negative), respectively.

The value of $\alpha$ is the probability of the critical region using
the parameter value defined by the null hypothesis. Conversely,
computation of $\beta$ is more complex, since $H_1$ is a composed
hypothesis rather than a single point hypothesis. To solve the problem
of defining the appropriate $\beta$, we considered the average of all
possible single hypotheses within the set of alternative
hypotheses. To perform this computation, we used a uniform prior for
the parameter of the alternative hypothesis, and considered the
predictive distribution for this prior choice. Fortunately, the
predictive is a uniform discrete distribution in the sample space and,
hence, is a constant equal to the inverse of the number of points of
the sample space. Therefore, the aforementioned average of $\beta$ is
the number of points within the acceptance region, divided by this
constant.

To choose the critical level, we considered, for a given value of y,
all possible critical regions.  The critical level is then the value
of $\alpha$ for the critical region that gives the smallest value of
each combination $a\alpha + b\beta$. We also defined an arbitrary
score S (eq. \ref{eq:score}), based on practical results, to order the
differentially expressed tags according to the relative ``distance''
between their corresponding \emph{p-value} and critical level.

\begin{equation}
  \label{eq:score}
  S=10\left(1-\frac{pv}{\alpha}\right)
\end{equation}

\subsection{Bayesian Significance Test: \emph{e-value}}
\label{sec:e-value}

The FBST, Full Bayesian Significance Test, was introduced by
\cite{Pereira1999}. The objective of this method is to obtain an
alternative index to \emph{p-values}, namely \emph{e-values} (where
{\bf e} stands for evidence). Both indices vary from zero to one,
since they represent probabilities on the sample space
(\emph{p-value}) and on the parameter space (\emph{e-value}). The
Bayesian test is defined on the original parameters $\vec
P_1,\cdots,\vec P_k$ of the libraries.  As a prior for these
parameters, we considered independent and identical Dirichlet
distributions with metaparameters $A_1,\cdots, A_M$.
Consequently, the posterior distribution for each independent
parameter $\vec P_j$ is Dirichlet with metaparameters
$x_{1j}+A_1,\cdots, x_{Mj}+A_M$
\citep{aitch_dir_logistic}. For this model, the posterior marginal
density of the parameter $p_{ij}$ is
$Beta(x_{ij}+A_i,N_j+S-x_{ij}-A_i)$, with
$S=A_1+\cdots+A_n$. From the independency of $P_j$, the
joint probability of all libraries, for each single tag, is a product
of these Beta densities.

To perform this test, one needs two numerical procedures: optimization
and integration. The low values of $x_{ij}+A_i$, compared to the
high values of $N_j+S-x_{ij}-A_i$, usually lead to an underflow
in the numerical integration. Thus, we transformed the Beta densities
into logistic-normal distributions \citep{aitch_dir_logistic}.  The
mean and the variance of the normal distributions, obtained after the
logistic transformation, $\zeta_{ij}=\log[p_{ij}/(1-p_{ij})]$, are
presented in eq. \ref{eq:mu} and eq. \ref{eq:sigma}, respectively,
where $\Psi$ is the digamma and and $\Psi'$ is the trigamma functions.

\begin{equation}
  \label{eq:mu}
\mu_{ij}=\Psi(x_{ij}+A_i)+\Psi(N_j-S-x_{ij}-A_i)
\end{equation}
\begin{equation}
  \label{eq:sigma}
  \sigma_{ij}^2=\Psi'(x_{ij}+A_i)-\Psi'(N_j-S-x_{ij}-A_i)
\end{equation}

Notice that the null hypothesis $\zeta_{i1}=\cdots=\zeta_{ik}$ is
equivalent to the original hypothesis $p_{i1}=\cdots=p_{ij}$. By using
this approach, the integration of the product of Beta distributions is
replaced by a integration of normal distributions, whose parameter
values avoid numerical representation problems that could arise when
using Beta distributions.

The Bayesian significance test was implemented on Basu, a C language
program named after Prof. Debabrata Basu, an Indian statistician who
motivated CABP to create the Full Bayesian Significance
Test \citep{Pereira1999}.

\section{Results}
\label{sec:final}

\subsection{The critical level as a function of y}

In order to establish an automatic procedure to discriminate tags
according to their differential expression status, we defined two sets
of weights for type I and type II errors, respectively: (a=1, b=1) and
(a=4, b=1). These values were arbitrarily defined based on several
analyses of SAGE (serial analysis of gene expression) data from human
libraries, followed by experimental validations with real-time PCR
(data not shown). The $\alpha$ critical levels were computed for a
range of possible values of y (the total tag frequency), and for k
values (number of libraries) ranging from 2 to 5. Since the
calculation of $\alpha$ is a computer intensive task, we estimated a
polynomial approximation of this critical level for each value of
k. This result was incorporated on Kemp, our implemented software for
the frequentist significance test, and is available in the
Appendix. Figures 1 and 2 present the dilog graphics of
the critical level values and the corresponding adjusted functions,
assuming k values of 2 and 5, respectively.

\begin{figure}[h]
  \centering
   \includegraphics[scale=0.55]{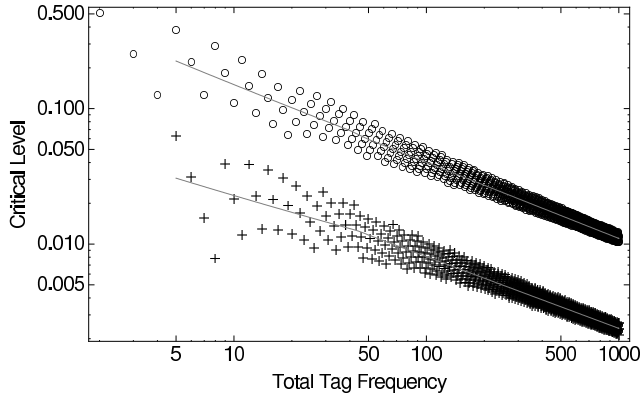}
   \caption{Dilog graphic with the simulated values of the critical
     level and the fitted function for k=2. Critical levels calculated
     with weights (a=1,b=1) are indicated by $\ocircle$, and with
     weights (a=4,b=1) are indicated by $+$.}
  \label{fig:cutoff2}
\end{figure}

\begin{figure}[h]
  \centering
   \includegraphics[scale=0.55]{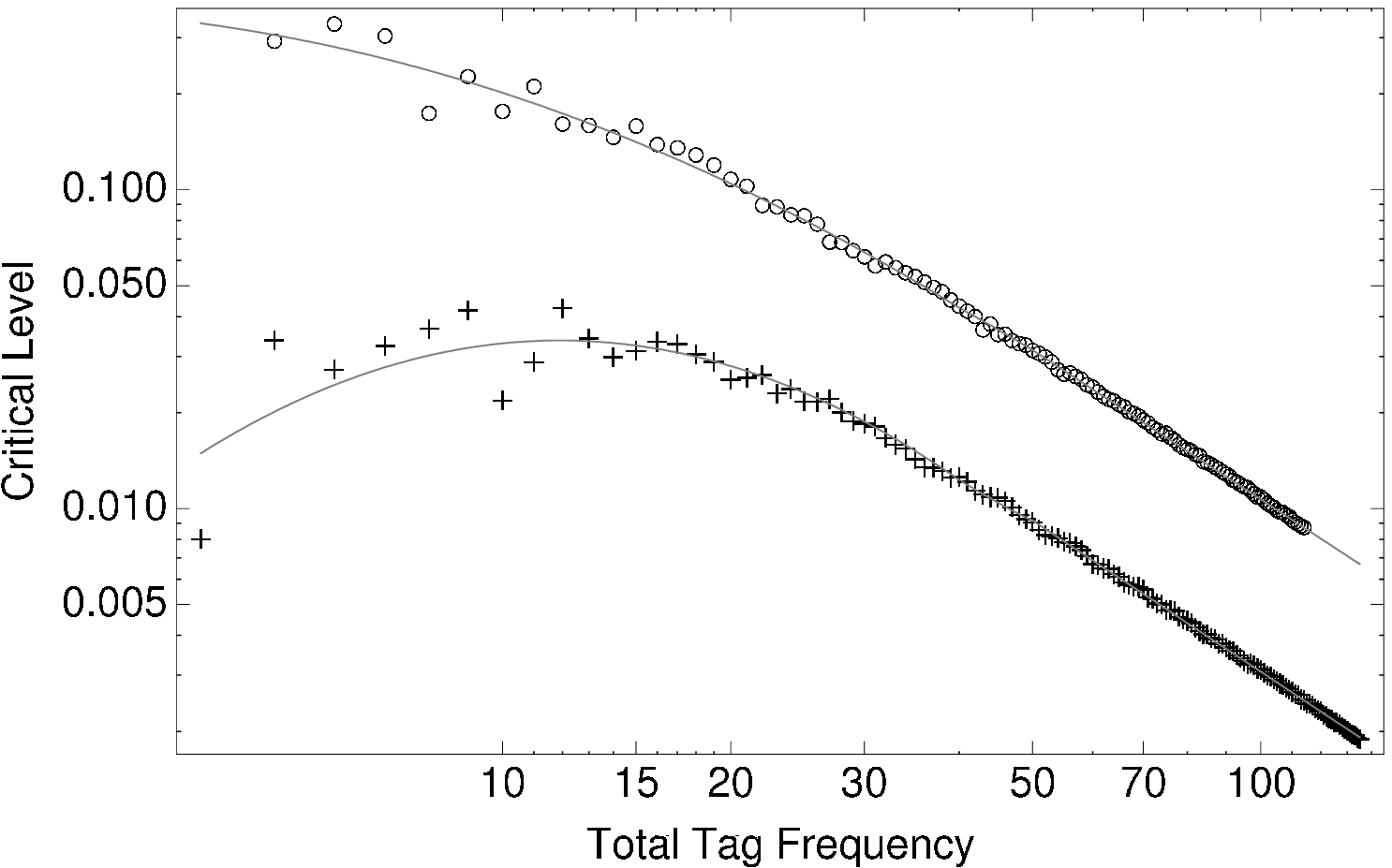}
   \caption{Dilog graphic with the simulated values of the critical
     level and the fitted functions for k=5. Critical levels calculated
     with weights (a=1,b=1) are indicated by $\ocircle$, and with
     weights (a=4,b=1) are indicated by $+$.}
  \label{fig:cutoff5}
\end{figure}

As can be seen, the set of weights (a=4,b=1) generates critical level
values that are consistently lower than those obtained using the set
(a=1,b=1). This result can be ascribed to weight 4 used for the
$\alpha$ error, which leads to a greater minimization of type I
error. Also, for both sets of weights, when y presents high values,
the critical level is much more stringent than the canonical values
0.1, 0.05 and 0.01.

\subsection{Comparison between Kemp \emph{p-value} and $\chi^2$ \emph{p-value}}
\label{sec:chi2}

\begin{figure}[h]
  \centering
   \includegraphics[scale=0.55]{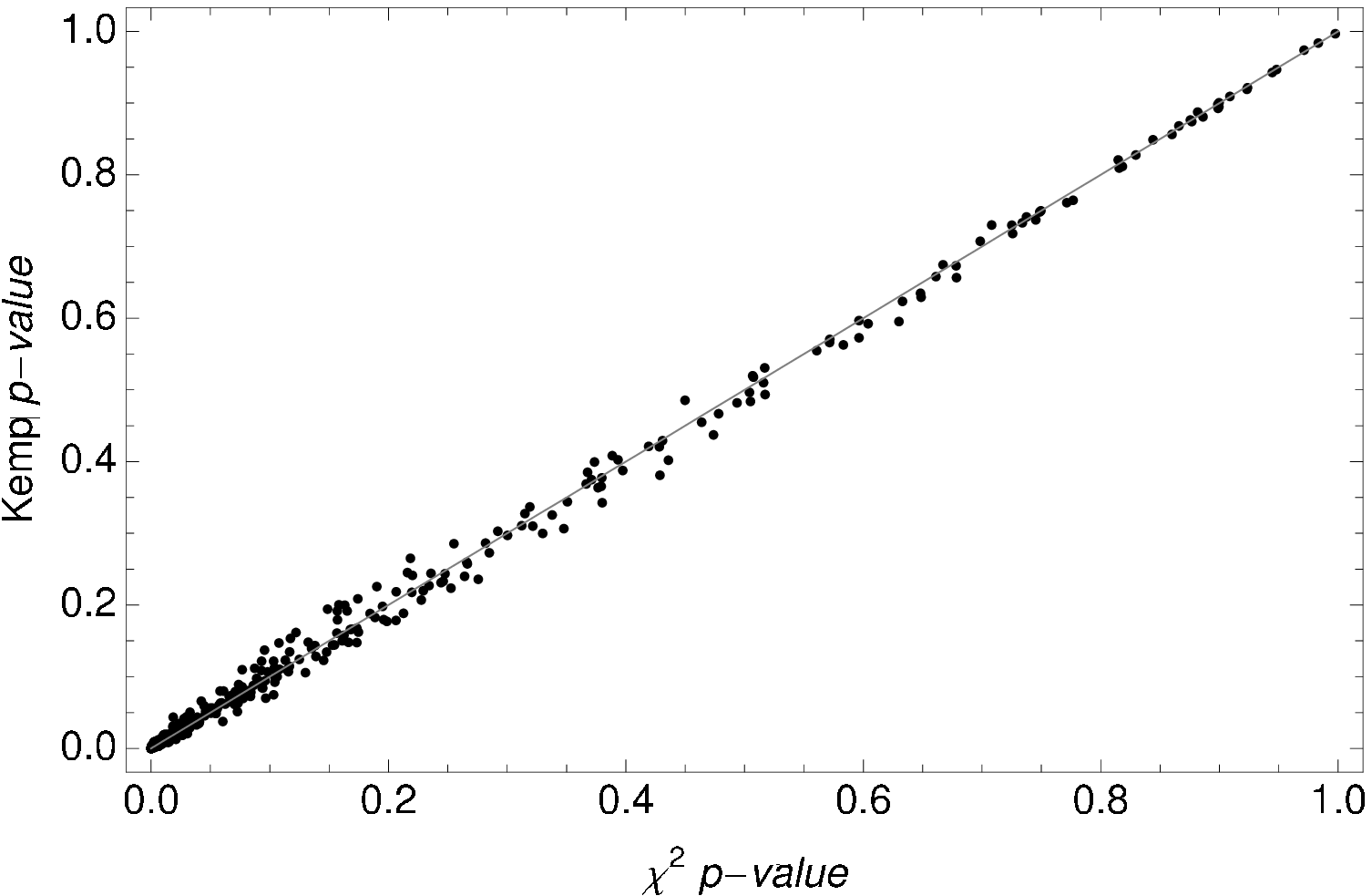}
   \caption{Relation between Kemp \emph{p-value} and $\chi^2$ \emph{p-value}
     for tags with $y>50$.}
  \label{fig:chi2-large}
\end{figure}

\begin{figure}[h]
  \centering
   \includegraphics[scale=0.55]{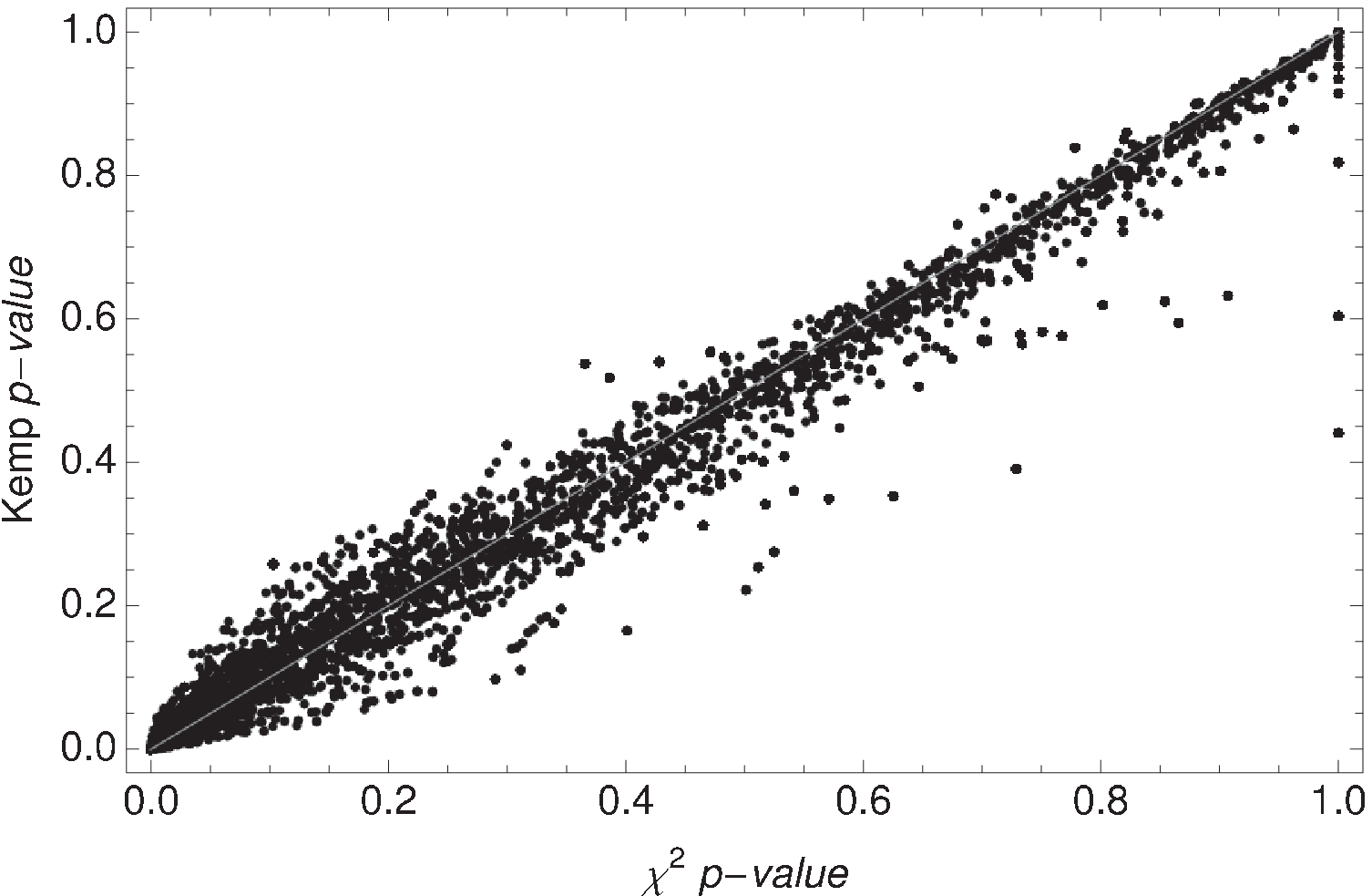}
   \caption{Relation between Kemp \emph{p-value} and $\chi^2$ \emph{p-value} for tags
     with $y\leq 50$}
  \label{fig:chi2-small}
\end{figure}

The $\chi^2$ homogeneity test is widely used and described in the
literature for comparison of digital expression profiles
\citep{man2000,romualdi2001}. We decided to compare our frequentist
significance level (\emph{p-value}) to the $\chi^2$ test. We used a
data set composed of four SAGE libraries derived from human brain
tissues, and potentially containing genes involved in increased risk
of Alzheimer’s Disease (GEO accession code GSE6677).  The tags were
arbitrarily separated into two groups according to their total
frequency ($y$). Thus, we calculated the significance levels using the
Kemp method and the $\chi^2$ test for both, the high-expression ($y
>50$, Fig. \ref{fig:chi2-large}) and low expression ($y \leq 50$,
Fig. \ref{fig:chi2-small}) tags. In Fig.  \ref{fig:chi2-large}, we can
see that there is a good agreement between the significance levels
when tags present a high expression level. Conversely,
Fig. \ref{fig:chi2-small} shows that when the expression of the tags
is relatively low, the levels present a much lower agreement. This
result is in consonance with what we should expect, since the $\chi^2$
test is asymptotic, whereas our proposed significance test is exact.

\subsection{Comparison between Kemp \emph{p-value} and Basu \emph{e-value}}
\label{sec:pvev}

\begin{figure}[h]
  \centering
   \includegraphics[scale=0.56]{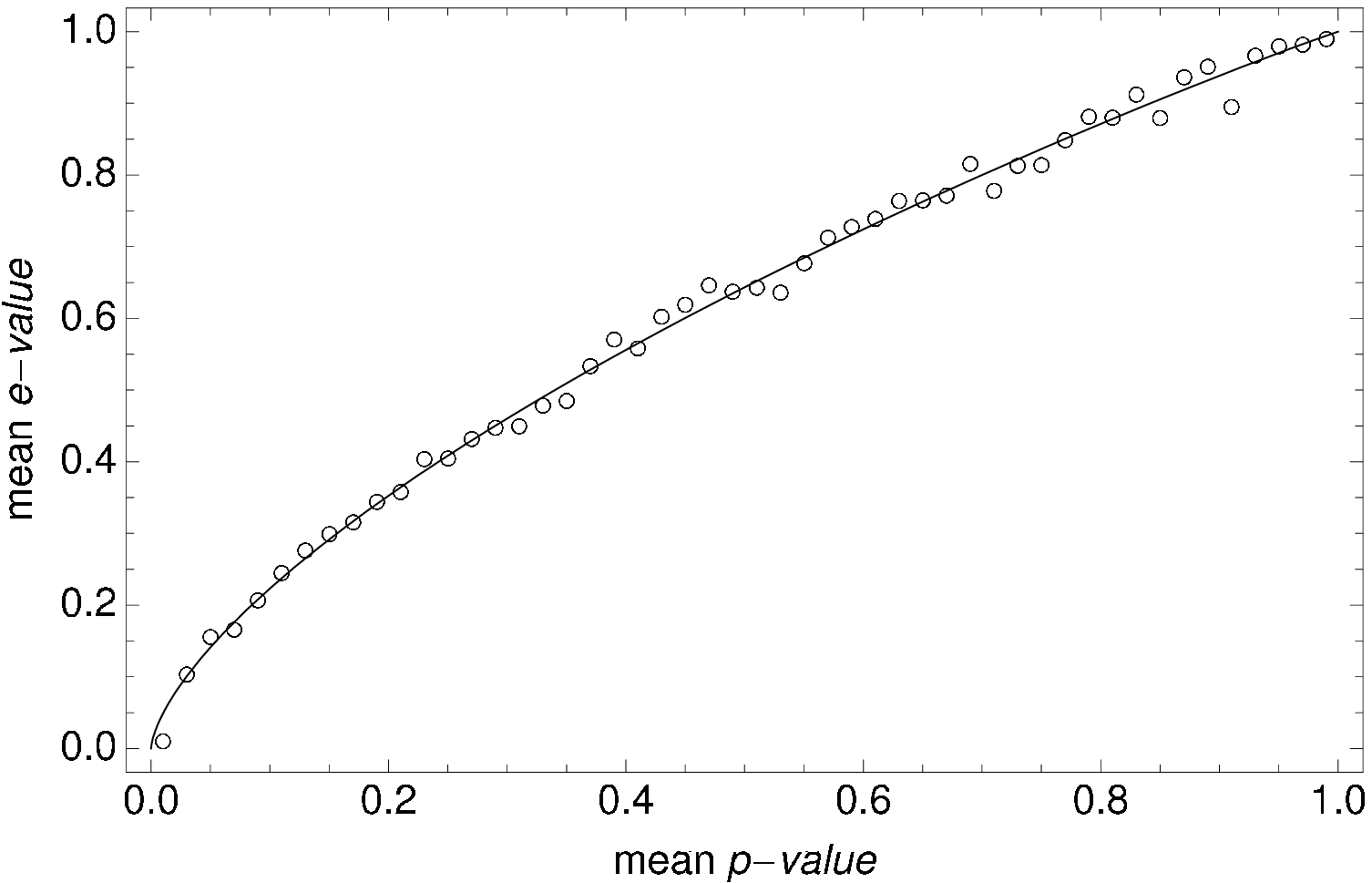}
   \caption{Relationship between \emph{p-value} and
     \emph{e-value}. $\ocircle$ indicate the pairs $(\bar p,\bar e)$
     and the curve the fitted function.}
  \label{fig:pvev-fit}
\end{figure}

Aiming at estimating the consistency of the proposed methods, we
compared the \emph{p-value} (the frequentist significance level), and
the \emph{e-value} (the Bayesian significance level) using the four
SAGE libraries described in the section \ref{sec:chi2}. To obtain the
relationship between \emph{p-values} and \emph{e-values} we calculated
the local weighted average of the \emph{e-values}, using the values of
$y$ as weights and a \emph{p-value} intervals of 0.04. This calculation
resulted into pairs $(\bar p, \bar e)$, where $\bar p$ is the center
of the interval, and $\bar e$ the weighted average. We then adjusted a
Beta distribution function to these pairs. The best fit obtained was a
Beta with parameters $a=0.66$ e $b=1.036$, corresponding to a mean of
0.39 and a standard deviation of 0.30. Fig. \ref{fig:pvev-fit}
displays a plot of the relationship between both significance levels,
and the fitted function. We can notice that there is a good agreement
between the \emph{p-value} and \emph{e-value}. Some differences
observed between their values can be explained by the fact that the
\emph{p-value} is an integral in the sample space, whereas the
\emph{e-value} is an integral in the parameter space. Despite this
difference, it is clear that both significance levels are most of
times convergent.

\section{Discussion}
\label{sec:discussion}

This paper introduces two novel methods for the comparison of digital
expression profiles. The methods, based on frequentist and Bayesian
statistics, were implemented on the open source programs Kemp and
Basu, respectively.  Several statistical tests have been used to
evaluate SAGE data and identify differentially expressed tags. Some of
these tests have been compared by different groups
\citep{man2000,romualdi2001,ruijter2002}. The general conclusion was
that the classical $\chi^2$ test, originally introduced by Karl
Pearson \citep{Pearson1900}, was equivalent to and even outperformed
other available tests, including the Fischer's Exact test, the test of
\cite{audic1997} and the R statistic of \cite{stekel2000}. The
$\chi^2$ test has the advantage of being simple and can be applied to
a broad range of problems. However, given the asymptotic character of
the $\chi^2$ test, it is not recommended for the analysis of low
frequency tags. Due to this feature, we conceived our frequentist test
using the original definition of more extreme sample points, without
any asymptotic result. As a consequence, our test is more correct than
the $\chi^2$ test for low expression tags. Corroborating this fact, a
comparison of the \emph{p-values} calculated by Kemp, and the $\chi^2$
\emph{p-value}, showed a good agreement (Fig. \ref{fig:chi2-large} and
Fig. \ref{fig:chi2-small}) only for high expression tags. For low
frequency tags, when the $\chi^2$ test becomes inappropriate, $\chi^2$
values showed a high disagreement with our calculated \emph{p-values}
(Fig. \ref{fig:chi2-small}). Thus, we believe that the frequentist
test proposed here, and implemented on Kemp, represents an improvement
for the analysis of digital expression profiles.

Some other methods \citep{baggerly2004,robinson2007,thygesen2006,
  vencio2004} have been proposed for the comparative analysis of SAGE
data. However, since these methods are designed for comparing groups
of libraries, their use is severely restricted in experiments where a
single library is represented in each category/condition.

The discrimination between high and low expression tags must be
performed in such a way as to consider both statistical and biological
relevance. Since housekeeping genes are expressed in high levels, the
absolute number of counts may present a considerable variation across
distinct libraries. These differences, nevertheless, are meaningless
from a biological standpoint. Conversely, some functionally important
genes present a relatively low expression, and exert their activity by
altering tiny amounts of their expression among the different tissues
and/or conditions \citep{wang2006}. Therefore, a tag presenting a
differential expression with low counts would not be considered as
significant by methods that use fixed critical levels, thus leading to
a misinterpretation of the data and loss of potentially valuable
information. This fact motivated us to calculate the critical level of
each particular tag taking into account its total frequency. If
population parameters are not exactly in the null sharp hypothesis set
(a set with a smaller dimension than the alternative hypothesis set),
highly expressed tags have smaller \emph{p-values} than low expression
tags. If one fixes the critical level, the minimum type I error, the
type II error decreases drastically when the sample size is
increased. Hence, it becomes difficult to accept a null hypothesis for
large sample sizes, and to reject it for small-sized samples. For
example, when comparing the expression of two 10,000-tag libraries, a
tag presenting counts of 7 and 21, respectively, would show a
\emph{p-value} of 0.013. Conversely, a tag with counts 10 and 30 would
result in a \emph{p-value} of 0.002. Considering a cutoff of 0.01, the
former tag would be considered as equally expressed in both libraries,
whereas the former tag would be interpreted as being differentially
expressed. This fact motivated us to calculate the critical level of
each particular tag as a function of the tag total frequency. The
critical region in our method is the one that minimizes a linear
combination of type I (the critical level $\alpha$) and type II ($\beta$)
errors. With this tag-customized approach, both tags of the example
above would be classified as differentially expressed, since their
corresponding critical levels would be 0.015 and 0.013,
respectively. The method is still coherent, since tags with very low
frequencies, even presenting differential counts, lead to high
significance levels. For instance, in the aforementioned example, tag
counts of 1 and 3 would result in a \emph{p-value} of 0.63, a much
higher value than the calculated cutoff of 0.03.  Concluding, our
method judges the tags in a fairer manner, since the cutoff value is
customized to any particular tag, according to its expression level.

The problem of significance testing of precise (sharp) hypotheses has
been controversial and both, the frequentist and Bayesian schools of
statistical inference, have offered solutions. As a counterpart to the
frequentist test introduced in this work, we decided to also offer an
alternative method, based on the previously described
\citep{pereira2007cst} Full Bayesian Significance Test (FBST). A clear
advantage of a Bayesian test, applied to digital expression profiling,
is that the total frequencies of the tags do not have to be fixed in
advance and are, in fact, unknown before the observation of the
libraries. This Bayesian procedure does not require any other
assumption in addition to the original multivariate Bernoulli
observations that produce the libraries. From a Bayesian standpoint,
it is important to judge hypotheses in their own environment, which is
the parameter space, not the sample space. With a more pragmatic view
on mind, we tried to check if the frequentist and Bayesian tests can
be related, even though they are defined in different spaces. In this
direction, we implemented the FBST in the Basu program, and compared
the \emph{e-values} to the \emph{p-values} previously determined by
Kemp (the frequentist test program), using SAGE libraries. To our
surprise, a strong correspondence between the averages of
\emph{p-values} and \emph{e-values} has been observed
(Fig. \ref{fig:pvev-fit}). This result indicates that both
methodologies can be reliably used to identify differentially
expressed tags. Also, because they lead to similar results using
totally different approaches, we believe that both methods can be used
in parallel to validate each other's results.

Kemp method fully implements a decision procedure, since it provides a
significance level and a critical level. Conversely, Basu method does
not calculate the most appropriate cutoff. For this task, one should
follow the decision theory steps described by \cite{pereira2007cst},
and build a loss function based on a good modeling of the risks
involved on deciding whether a tag should or not be considered as
differentially expressed.  In this direction, our group is currently
working on the development of a critical level for the
FBST. Alternatively, one can use an approximation to determine a
cutoff for Basu. Since the \emph{p-values} and \emph{e-values} are
linked by a Beta distribution function, as we have shown for a set of
SAGE libraries from human brain tissue (Fig. \ref{fig:pvev-fit}) and
some other datasets (data not shown), we propose to use the Kemp
cutoff, properly adjusted by the linking function.

Concluding, the frequentist and Bayesian significance tests reported
in the present work, and implemented in standalone open-source
programs, extend the set of currently available statistical tests for
digital expression profiles. Also, we believe that they offer some
advantages over other reported tests, including a more adequate
treatment of low expression tags and the automatic calculation of a
customized critical level.

\section*{System Requirements}

The source code of KempBasu package and a executable binary for MS
Windows are publicly available at the address
\url{http://code.google.com/p/kempbasu/}, and are distributed under the
GNU General Public License. The code depends on glib, GSL and Judy
libraries, and if the Pthreads API is available, KempBasu can be run
using multiple processors. Tested platforms include Linux, MacOSX and
MS Windows.

\section*{Acknowledgments }
LV received a PhD fellowship from CAPES. AG and CABP are research
fellows of CNPq.

\bibliography{kempbasu}

\begin{thebibliography}{}
\bibitem[Pearson(1900)Pearson]{Pearson1900}
Pearson, K. (1900).
\newblock {On the criterion that a given system of deviations from the probable
  in the case of a correlated system of variables is such that it can be
  reasonable supposed to have arisen from random sampling}.
\newblock {\em Philos. Mag.\/}, {\bf 50}, 157--175.
\end{thebibliography}

\newpage 

\section*{Appendix}

To establish the function of y that gives the approximate $\alpha$, we
consider the pairs $\{L=Log(y);Log(\alpha)\}$ and use the least
squares method piecewisely in two difference regions of $y$ values:
[1;50] and [51;10,000] . The former region adjusts a second degree
polynomial $$\alpha_k^{(1)}(y)=a_kL^2+b_kL+c_k$$ and the latter region
adjusts a line
$$\alpha_k^{(2)}(y)=u_k+v_kL.$$ Tables \ref{tab:coeffs} and \ref{tab:coeffs11} present
the coefficient values for those functions for weights $(a=4,b=1)$ and
$(a=1,b=1)$, respectively . The linear and the quadratic functions are
combined, for each value of $k$, into a single continuos function by
the eq. \ref{eq:alpha-piecewise}. 

\begin{equation}
  \label{eq:alpha-piecewise}
  \alpha_k(y)=\left\{
    \begin{array}{ll}
      \alpha_k^{(1)}(y) & \text{para } y<40\\
      (1-\lambda)\alpha_k^{(1)}(y)+\lambda\alpha_K^{(2)}(y) & \text{para } 40\leq y<50\\
      \alpha_k^{(2)}(y) & \text{para } y\geq50 \\
    \end{array}
    \right.
\end{equation}

\begin{table}[h]
  \centering
  \begin{tabular}{lrrrrr}
\hline
k        &$a_k$	        &$b_k$	        &$c_k$	        &$u_k$	        &$v_k$\\
\hline
2	&0.00957978	&-0.463118	&-2.76474	&-2.37781	&-0.530119\\
3	&-0.304365	&1.18976	&-4.60784	&-0.713611	&-0.968513\\
4	&-0.931159	&5.00318	&-10.1863	&0.385118	&-1.28105\\
5	&-0.685327	&3.39467	&-7.59502	&1.47602	&-1.57657\\
6	&-0.914225	&4.84175	&-9.81444	&1.93518	&-1.70783\\
\hline    
  \end{tabular}
  \caption{Coefficient values of fitted critical level functions for the minimization of $4\alpha+\beta$}
  \label{tab:coeffs}
\end{table}

\begin{table}[h]
  \centering
  \begin{tabular}{lrrrrr}
\hline
k        &$a_k$	        &$b_k$	        &$c_k$	        &$u_k$	        &$v_k$\\
\hline
2	&0.00748022	&-0.607463	&-0.53588	&-0.629139	&-0.561742\\
3	&-0.226299	&0.503742	&-1.7504	&0.677628	&-0.968169\\
4	&-0.215143	&0.334093	&-1.38061	&1.79399 	&-1.30545\\
5	&-0.248689	&0.369967	&-1.13529	&2.62984	&-1.55664\\
\hline    
  \end{tabular}
  \caption{Coefficient values of fitted critical level functions for the minimization of $\alpha+\beta$}
  \label{tab:coeffs11}
\end{table}

\end{document}